# A TTP by TTP Approach: Precise Malware Detection via Malicious TTP Recognition

Yashovardhan Sharma
*University of Oxford*
*Oxford, United Kingdom*
*yashovardhan.sharma@cs.ox.ac.uk*

***Abstract*—Machine learning methods, and especially neural networks, are now routinely used for malware detection in network traffic. Though very effective, systems based on such methods often $(i)$ are purely data-driven, ignoring the substantial body of available knowledge about the tactics, techniques, and procedures (TTPs) possibly used, and, consequently $(ii)$ are not precise, since they either cannot correlate malicious activity with TTP usage, or if they do, they are unable to explain which TTP has been maliciously used. In this paper we demonstrate that it is possible to precisely detect malware by $(i)$ providing the neural network model with information about the TTPs used by any given sample, and $(ii)$ teaching the neural network to detect not just the malicious activity as a whole, but which specific TTPs are maliciously used. We show that our approach consistently outperforms the three alternative models, which either do not exploit TTP information, or which are not taught to detect the malicious usage of TTPs, or both. Moreover, we show that our approach $(i)$ is particularly beneficial in detecting malware that utilises rarely-used TTPs, a scenario which is particularly challenging for the other systems; $(ii)$ allows for TTP by TTP tuning, further improving its ability to detect the malicious usage of TTPs; $(iii)$ consistently outperforms other systems across a wide-range of scenarios, including when relying on limited training data or when subjected to adversarial attack.**

## 1. Introduction

Machine learning methods and especially neural networks are now routinely used for malware detection in network traffic (see, e.g., [18], [20], [32]). Just within the research literature, Scopus indicates that there has been a 508% increase in papers referencing machine learning (ML) or deep learning (DL) within the context of cyber security over the last 5 years [36]. The rapid increase in the usage of such techniques within the domain of malware detection has been fuelled by the challenge of keeping up with constantly-evolving malware and the ability of ML models to handle large volumes of data. Approaches using neural networks in particular have been favoured since they do not require an expert's domain knowledge to define discriminative feature used in malware detection system [9]. Though very effective, such methods are often: $(i)$ purely data-driven, ignoring the substantial body of available knowledge about the tactics, techniques, and procedures (TTPs) possibly used, and $(ii)$ not precise, since they either cannot correlate malicious activity with TTP usage or if they do, they give little to no explanation about which TTPs have been maliciously used. Furthermore, ML methods typically require large amounts of quality training data, and are often not robust to adversarial attacks [9], [32].

In this paper, we leverage the ontology of adversarial behaviour provided by the MITRE ATT&CK framework, comprising of tactics, techniques, and procedures (TTPs), to show that it is possible to precisely detect malware from network captures by $(i)$ formulating the problem as a multi-label classification problem (see, e.g., [43]) with one set of labels to classify the TTPs, and another label to classify the maliciousness of the entire network traffic sample (henceforth referred to as *sample*), and $(ii)$ designing a Deep Neural Network (DNN) architecture for the multi-label classification problem, that can take into account information about the TTPs utilised by the sample in the form of TTP features. In particular, we propose a methodology in which $(i)$ each sample is analysed in order to automatically extract the TTPs utilised by it, $(ii)$ the network traffic features of the sample, together with the features about the matched TTPs, are then given as inputs to the DNN architecture, which is not only responsible for detecting the maliciousness of the sample, but also which TTPs are maliciously used. To the best of our knowledge, ours is the first system which is able to detect malicious network traffic and precisely characterise which TTPs have been maliciously used. This is a rather useful property since it enables our system to have different fine-grained outputs, as advocated in [3]. Compared to the existing purely data-driven approaches for malware detection (see, e.g., [18], [20]), we expect that by $(i)$ providing the information about the TTPs used by a sample, and $(ii)$ training the neural network to detect both the malicious activity as a whole and the specific TTPs that are maliciously used, our system is particularly well-suited for detecting malware that utilises uncommon or rarely-used TTPs—a scenario which is particularly challenging for any data-driven approach.

We evaluate our system by conducting an extensive experimental analysis involving more than 1.5 million malicious and benign samples, curated from 5 different publicly available datasets, thus allowing us to test our system on a broad range of real-world samples. We consider 5 micro-average metrics usually used to evaluate systems for multi-label classification problems ($F_1$-score, Subset Accuracy, Precision, Recall and False Positive Rate), together with the macro-average metrics corresponding to the $F_1$-score, Precision and Recall. These metrics are widely used (see, e.g., [8], [24], [33]), and each highlights a different aspect of our system's performance. For instance, we utilise the macro-average metrics to highlight the performance of our system at TTP classification since individual TTP classes are unbalanced in our data, and these metrics are suited for multi-label classification with unbalanced classes [30], [42]. On the other hand, we use micro-average metrics to highlight the overall performance of our system, specifically for sample classification, which is evaluated on balanced malicious and benign datasets. We further evaluate our approach across four possible configurations:

1) STTP2STTP: the system is given the sample and TTP features as input, and it has to detect both if the sample is malicious, and which TTPs are maliciously used by it.
2) STTP2S: the system is given the sample and TTP features as input, and it has to determine whether the sample is malicious or not.
3) S2STTP: the system is given just the sample as input, and it has to detect both if the sample is malicious, and which TTPs are maliciously used by it.
4) S2S: the system is given just the sample as input, and it has to determine whether the sample is malicious or not.

We test our approach in all configurations, across all of our datasets, to thoroughly evaluate its performance. We conduct further experimental analysis to test our approach in various challenging conditions, such as when faced with limited training data or dealing with adversarial attack. Our experimental analysis confirms that:

- The STTP2STTP, STTP2S and S2STTP models consistently outperform the standard S2S model.
- The STTP2STTP model has the best overall performance, and significantly better average performance in detecting rarely-occurring malicious TTPs, as witnessed by an average improvement in the Macro $F_1$-score, Macro Precision and Macro Recall of 37.55%, 11.82% and 46.05% respectively.
- It is possible to further improve the performance of STTP2STTP by finely tuning the decision threshold for each individual TTP. In our datasets, we measure average performance gains of 12.5%, 12.0%, and 47.6%, for the Macro $F_1$-score, Macro Precision, and Macro Recall respectively.
- STTP2STTP maintains its ability to better detect malicious TTPs than S2STTP even when considering challenging scenarios where models are trained with a limited number of samples, or when tested with noisy malicious samples flooded with benign traffic by adversaries attempting to camouflage their activities.

The better performance of STTP2STTP when compared to any other model (be it S2STTP, STTP2Sor S2S) corresponds to the possibility of training STTP2STTP with less data and still getting the same performance of the other model, thus at least partially overcoming the well-known data-greediness problem of ML models (see, e.g., [9]). For instance, STTP2STTP utilising 25% of the training data has a higher average $F_1$-score than S2STTP utilising 100% of the training data, and STTP2STTP when trained with 10% of the data has on average still a higher Recall than S2STTP trained with 100% of the data.

To recap, the main contributions of the paper can be summarised as follows:

- We design a novel multi-label DNN architecture that can precisely detect network-based malware along with the specific TTPs utilised by it, by exploiting TTP information in the form of automatically-generated features.
- We outperform the state-of-the-art models, regardless of whether or not they exploit the same TTP features used by our system. Our approach is particularly beneficial in detecting the malicious usage of uncommon or rarely-used TTPs.
- Our system is highly configurable, allowing for TTP by TTP tuning, thus further increasing its performance.
- We test our approach on a large dataset of real-world malicious and benign samples. We utilise 5 balanced datasets to fairly test our system, consisting of over 1.5 million samples.
- We evaluate our system under challenging scenarios, such as when it has access to limited training data or when it is subjected to adversarial attack, and demonstrate that it consistently outperforms other systems even under these conditions.

The rest of the paper is organised as follows: Section 2 gives a brief overview of the MITRE ATT&CK framework and discusses closely related work. Section 3 discusses the overall architecture of the system and the functioning of its individual components. The four models STTP2STTP, S2STTP, STTP2S and S2S are then comparatively evaluated in Section 4. Section 5 contains further experimental analysis of our system, by testing it under various challenging conditions and tuning it on a TTP-by-TTP basis to further improve performance. A thorough discussion about the results presented is in Section 6, followed by the overall conclusions of our work in Section 7.

## 2. Background and Related Work

The MITRE ATT&CK framework [35] is a well-known publicly available knowledge base of adversarial tactics, techniques and procedures, based on real-world observations. Within the ATT&CK framework:

- A *tactic* represents a tactical goal of the adversary, i.e., the reason why the adversary performs any action.
- A *technique* is the method by which the adversary achieves their tactical goal. Techniques can have *sub-techniques*, corresponding to a more specific action within this method.
- A *procedure* is a specific implementation of the method the adversary uses to fulfil their tactical goal.

As an example, an adversary may dump OS credentials (*technique* T1003) to achieve credential access (*tactic*). This may be done by dumping the contents of `/etc/passwd` or `/etc/shadow` (*sub-technique*

T1003.008), using a tool such as LaZagne[1] (*procedure*). The MITRE ATT&CK framework is frequently updated to keep it up-to-date with the evolution of malware and novel adversarial approaches. The current version of the ATT&CK framework for Enterprise features 14 tactics, 201 techniques, and 424 sub-techniques. This paper will focus specifically on the techniques (and encompassing tactics) that utilise the network to achieve their goals, which is a smaller subset of the overall techniques present in the ATT&CK framework.

The ATT&CK framework and the TTPs described within it have been used extensively to tackle various challenges within the context of cyber security. The broad idea is to exploit existing human domain knowledge in an automated manner, in order to assist solving problems that rely on the analysis of adversarial behaviour. TTPs from the ATT&CK framework have been utilised to track the evolution of malware and describe emerging trends [4], [23], locate TTPs within the control flow graph of malware executables [7], automatically extract TTPs from Cyber Threat Intelligence (CTI) reports [14], [19], [22], discover existing inter-dependencies in a TTP chain [1], detect Living-Off-the-Land (LotL) malware techniques and Advanced Persistent Threat (APT) attack campaigns [12], [17], and build malware detection systems that exploit TTP information [29]–[31].

If we focus specifically on how TTPs have been used for the automatic detection of malicious activity, to the best of our knowledge, the only research literature that has explored this direction are the recent works by Sharma *et al.* [28]–[31]. In [30] and [29], the authors create an explainable and extensible IDS that exploits TTPs in order to detect malware. In this system, a different dataset is created for each TTP, where the label is considered positive if the sample is malicious and the TTP is present. For each dataset, a different decision tree [41] is trained and then used together with the others as an ensemble to decide on the maliciousness of the sample, thus resulting in an extensible and explainable system. However, the limitation of such an approach is that its modularity is achieved by having a separate ML engine for each TTP, thus sacrificing the possibility of exploiting the existing correlations between different TTPs to detect malicious activity (correlations among labels are often exploited to achieve better performance, see, e.g., [25], [34]). Additionally, this system conducts its analysis on a per-flow basis, which while effective, is unable to consider all the flows in a sample together and thus cannot evaluate the behaviour of a malware sample as a whole. On the other hand, in [31] the authors create binary classification models, called TTPxML models, which incorporate TTP information as features into these models in order to detect network-based malware. They test this approach on a wide-variety of ML models, such as Support Vector Machines [5], Random Forests [13], and DNNs [21], and find that TTP features can be exploited to improve the malware detection capabilities of ML-based systems. Here, TTPs are first automatically extracted from the network sample, then a vectorial representation is created of both the sample features and the TTP features, which is subsequently provided as input to a binary classifier which labels the entire sample as malicious or not. This system is the state-of-the-art in terms of exploiting TTP information in an automated manner to detect network-based malware. Thus, we closely model this system in our STTP2S model, in order to be able to compare and contrast the results of our multi-label classification-based approach with their binary classification-based system.

1. https://github.com/AlessandroZ/LaZagne

# 3. System Architecture

At a high level, our system classifies a given sample in the following manner:

1) A raw network traffic sample is taken as input, and modelled into a vectorial representation of the network traffic data, known as Bag of Flows (BoF).
2) TTP information about the sample is automatically extracted, and incorporated into the Bag of Flows in the form of TTP features.
3) This Bag of Flows is given as input to our multi-label DNN architecture, which performs the malware classification.
4) As per the model configuration, our system gives as output: $(i)$ a malicious or benign classification for the overall sample, or $(ii)$ a multi-label classification of each maliciously-used TTP by the sample, in addition to the malicious or benign classification for the overall sample.

We broadly describe the specifics of each of these steps in the subsequent subsections.

## 3.1. Constructing the Bag of Flows

The first two steps are largely based on the methodology of the TTPxML system described in [31]. Here, to make the paper more self-contained, we briefly describe how the raw network traffic samples are converted into their Bag of Flow representation, as required by the next step in our malware detection system. See [31] for more details.

Each raw network traffic sample is represented as a set of network flows, using Yet Another Flowmeter (YAF) [15], and annotated with the set of features in Table 1. YAF has several advantages, including the fact that it has been designed for high performance. Then, each flow is annotated with the 13 TTPs from the MITRE ATT&CK framework as shown in Table 2. This TTP matching is done at the flow level, and utilises the methodology described in [30]. These 13 TTPs encompass 11 tactics, 9 techniques, and 6 sub-techniques, and thus represent a sufficiently wide variety of adversarial behaviour that can be exhibited by network-based malware. Each flow is then converted into a vector in which $(i)$ each non-numerical feature is converted into a numerical one, $(ii)$ the categorical features “End-reason”, “Protocol” and the 4 features representing TCP flags, are represented as one-hot encoded vectors, while $(iii)$ “Destination IP” is converted into a 32-bit integer maintaining the structure information of IP addresses. Any flow properties that are not useful for classification or would unintentionally bias it, are dropped. Finally, the remaining flow properties are aggregated in a Bag of Flows, which is a vectorial representation of the original sample. Each BoF is constructed by $(i)$ min-max scaling each flow-level feature, and $(ii)$ transforming each flow-level feature into a normalised histogram recording the proportion of values that fall within a certain range. In practice, for each non categorical feature $B = 5$ bins are defined, and for each bin, the proportion of flows in the sample having a value belonging to that bin is recorded.

TABLE 1: Flow properties.

| Field | Property |
| --- | --- |
| Flow Hash | The hash associated with each unique flow. |
| Sample Hash | Hash of the sample corresponding to the flow. |
| Unique ID | A unique ID associated with each flow. |
| Dataset | The dataset to which a flow belongs. |
| Source Port | The port from which the flow originates. |
| Destination Port | The port on which the flow connects to. |
| Start time, end time | Flow start or end time. |
| Duration | Flow duration in seconds. |
| Protocol | IP protocol identifier in decimal format. |
| Entropy | The Shannon Entropy for the flow payload. |
| Applabel | The application label, as identified by YAF. |
| Source IP, Dest. IP | Source and destination IPv4/IPv6 address. |
| Type, Code | ICMP type or code in decimal format. |
| Isn, Risn | Forward or reverse TCP sequence number. |
| Flags | 4 properties representing various TCP flags. |
| Tag, Rtag | 802.1q VLAN tag in forward/reverse direction. |
| Pkt, Rpkt | No. of packets in forward/reverse direction. |
| Oct, Roct | No. of bytes in forward/reverse direction. |
| RTT | Round-trip time estimate in milliseconds. |
| End-reason | Reason for termination of flow. |

In the STTP2S and STTP2STTP models, the Bag of Flows representation also includes the following information about the TTPs matched by the flows in the sample:

- For each TTP two aggregate features are created: one representing the proportion of flows in the sample that match the given TTP, and the other representing the total number of flows matching the given TTP. This last feature is normalised over all the samples.
- Then, the following information about the matched TTPs is also provided: $(i)$ the number of unique TTPs matched by the sample, $(ii)$ the total number of TTPs matched by the sample, $(iii)$ the number of flows matching at least one TTP in the sample, $(iv)$ the proportion of flows matching at least one TTP in the sample, $(v)$ the mean of the number of TTPs matched by the sample, and $(vi)$ the standard deviation of the number of TTPs matched by the sample. All these features are normalised over all the samples.

At this point, the sample has been successfully modelled as a Bag of Flows, and its corresponding TTP information has also been incorporated into this BoF. This vectorial representation of the given sample is then passed on to the classifier.

## 3.2. Multi-label DNN Classification

The multi-label DNN architecture utilised by our system has an input layer of neurons, that has the same size as that of the vector representing the Bag of Flows, which was created in the previous step. The neural network is implemented as a fully-connected feed-forward neural network with ReLU non-linearity and four hidden layers. The output layer of the DNN contains 14 neurons, 13 corresponding to the outcome for the 13 TTPs supported by our system, and the final neuron corresponding to the maliciousness of the entire sample. Thus, this architecture allows us to operate in both multi-label and binary settings, enabling us to test our system in various configurations.

TABLE 2: List of supported TTPs from the MITRE ATT&CK framework.

| Tactics | Techniques |
| --- | --- |
| **Reconnaissance** | T1590 - Gathering Victim Network Information |
| **Credential Access** | T1557.001 - Man-in-the-Middle |
| **Discovery** | T1124 - System Time Discovery<br>T1135 - Network Share Discovery |
| **Lateral Movement** | T1021.001 - Remote Services (RDP)<br>T1021.004 - Remote Services (SSH)<br>T1550.003 - Use Alternate Authentication Material<br>T1563.001 - Remote Service Session Hijacking (RDP)<br>T1563.002 - Remote Service Session Hijacking (SSH)<br>T1570 - Lateral Tool Transfer |
| **Command and Control** | T1071 - Application Layer Protocol<br>T1090 - Proxy<br>T1105 - Ingress Tool Transfer<br>T1571 - Non-Standard Port |
| **Execution** | T1053 - Scheduled Task/Job |

As previously stated, our system allows for 4 different configurations depending on whether:

1) TTP features are given as input along with the sample's BoF or not, to the classification system.
2) The output of binary classification of the sample, should also include information about which TTPs are maliciously used by the sample or not.

These 4 configurations are: STTP2STTP (TTP information in both input and output), S2STTP (just the sample information in input and TTP information as output), STTP2S (TTP information in input and just the sample maliciousness as output), and S2S (just the sample information in input and sample maliciousness as output). All the four classification tasks can be modelled as multi-label classification problems, and transformed into a binary classification problems, when dealing with configurations where the system only has to determine whether the overall sample is malicious or benign. In a multi-label classification model, an input vector $x$ of values is mapped to an output binary vector in which each element represents a property of the input sample $x$. In contrast to binary classification, an input sample $x$ can have multiple elements in the output vector mapped to 1, thus corresponding to $x$ satisfying the properties represented by such output values. Formally, a *multi-label classification* problem is defined as a pair $(\mathcal{A}, \mathcal{X})$, where $\mathcal{A}$ is a set of labels, while $\mathcal{X}$ is a finite set of pairs $(x, y)$ where $x \in \mathbb{R}^N$ $(N \geq 1)$ is a data point/sample, and $y \subseteq \mathcal{A}$ is the ground truth of $x$, i.e., the set of labels associated with $x$. A *model* $m$ for $\mathcal{P}$ is a function $m(\cdot, \cdot)$ which maps every label $A$ and data point $x \in \mathbb{R}^N$ to $[0, 1]$. For every label $A$ the function $m_A : \mathbb{R}^N \mapsto [0, 1]$ is defined by $x \mapsto m(A, x)$, for each data point $x \in \mathbb{R}^N$. Finally, a data point $x \in \mathbb{R}^N$ is predicted by $m$ to have label $A$ whenever $m_A(x)$ is greater than or equal to a user-defined threshold $\theta_A \in [0, 1]$.

Ideally, we would like to have one label per TTP that we are interested in detecting. In our case, this would correspond to having the set of labels $\mathcal{A}$ equal to the set of TTPs in Table 2; the set $\mathcal{X}$ to be the set of Bag of Flows corresponding to the samples, in which each Bag of Flows $x$ is paired with the labels $y$ corresponding to the TTPs in Table 2 which are maliciously used by the sample. Such a picture has two difficulties:

1) All available datasets of malware network traffic only have information regarding the maliciousness of the sample, and do not contain any information about which are the TTPs being used.
2) The list of TTPs in Table 2 is by no means exhaustive, meaning that there can certainly be malicious network traffic samples that do not match any of the TTPs supported by our system.

We overcome the first difficulty by $(i)$ automatically detecting the TTPs in each sample as described in the previous subsection, and $(ii)$ associating the set of labels corresponding to the detected TTPs, to the sample.

To overcome the second difficulty, we introduce an additional label called Mal-Sample whose ground truth is assumed to be 1 when the sample is malicious, even if it does not match any of the TTPs our system supports. Notice that even assuming that our system would support all known TTPs, such an additional label is still useful given the possibility of either an attack using a never seen before TTP, or the TTP detection component failing to recognise a TTP (be it malicious or benign) in the network traffic sample. However, the presence of the label Mal-Sample indicating sample maliciousness and of the other labels indicating malicious usage of TTPs raises the following possibilities:

1) The TTP outputs and the sample output are compatible with each other, meaning that either all the TTPs in the sample and the sample itself are classified as benign, or that some of the TTPs in the sample and the sample itself are classified as malicious. In this case, no issue arises and all the TTPs and the sample are considered according to their predictions.
2) The TTP outputs and the sample output are different because all the TTPs in the sample are labelled as benign and the sample as malicious. Such a case reflects the possibility that the model has detected a malicious sample even though it does not match any of the supported TTPs. Given this, all the TTPs and the sample are considered according to their predictions, as in the first case.
3) The TTP outputs and the sample output are different, implying that some TTP is labelled as malicious while the sample is labelled as benign. Such a situation does not reflect any real scenario, and it is possible given the inherent fact that machine learning models in general, and neural networks in particular, may fail to learn existing relationships between the labels (see, e.g., [10], for a discussion on this). In such a case, taking a prudential approach, we also consider the sample as malicious and count it accordingly in the metrics.

TABLE 3: Statistics about the original datasets.

| Dataset | Samples | Flows |
|---|---|---|
| Ember Malicious [2] | 435,741 | 26,567,527 |
| MalRec [27] | 37,763 | 12,273,502 |
| MalShare [37] | 1,268,923 | 32,310,907 |
| VirusShare [38] | 595,098 | 12,217,493 |
| Ember Benign [2] | 155,432 | 1,423,023 |
| Total (unique) | 2,131,832 | 83,371,480 |

Indeed, from a technical point of view, the problem we are considering is the special case of a multi-label classification problem with hierarchical constraints between the labels. Special techniques have been developed for such problems in order to guarantee coherency between hierarchical labels, (see, e.g., [11], [26], [40]). We did not consider such techniques given the relative simplicity of our hierarchy with only one level, in which, as we said, coherency is guaranteed by simply flipping the Mal-Sample label when necessary. In practice, such incoherence shows up very rarely: in our experiments there are no differences (up to the third decimal place) in the $F_1$-score, Precision, and Recall of label Mal-Sample when measured before and after Mal-Sample is turned to 1, in order to solve the incoherence between label Mal-Sample and the TTP labels.

# 4. Evaluation

The goal of our experimental evaluation is to assess the benefits of $(i)$ providing TTP information as additional features to the neural network; and $(ii)$ teaching the neural network to detect the malicious usage of TTPs, when classifying a given sample. Specifically, we pose the following research questions:

1) Is it possible to effectively detect the malicious usage of each TTP?
2) In the previous case, do the TTP features help?
3) Does detecting the malicious usage of TTPs, also help solve the simpler problem of determining the maliciousness of the overall sample?
4) In the previous case, do the TTP features help?

Towards this end, we comparatively evaluate our different models STTP2STTP, S2STTP, STTP2S and S2S on the same 5 datasets used in [31]. In the following subsections, we first provide a detailed description of the datasets, highlighting some of their characteristics. Then, we present the metrics used in our evaluation. Finally, we present the results of our experimental evaluation, addressing the first two questions in Subsection 4.3, and the third and fourth questions in Subsection 4.4.

## 4.1. Datasets

The datasets consist of a varied population of malware and benign samples from the five publicly available datasets represented in Table 3. Of these five datasets, only one, namely "Ember Benign", contains benign samples. Starting from the datasets in Table 3, five new datasets have been constructed by randomly sampling the five original ones. Each of the five new datasets has then been split into training (60% of the samples), validation (20% of the samples), and test (20% of the samples) sets. These new datasets share the following properties: $(i)$ they are balanced at the sample level, meaning that they contain a balanced number of malicious and benign samples, $(ii)$ they are split ensuring that they have pairwise disjoint test sets, $(iii)$ they are as large as possible given the previous two requirements. Though these new datasets should share the same distributions of samples, by considering five (and not just a single dataset) we are able to compute the mean and standard deviation of our results, thus minimising the risk of skewed results due to particular distributions in the training, validation, or test sets.

TABLE 4: Statistics about the newly-derived datasets. Average number and percentage of matched samples per TTP, across the train/validation/test sets. Column #s is the number of matched samples. The Total number of TTPs does not include Tnull, since Tnull represents the samples that do not match any TTPs.

| | **Train** (tot. samples = 186183) | | | | **Validation** (tot. samples = 62060) | | | | **Test** (tot. samples = 62061) | | | |
|---|---|---|---|---|---|---|---|---|---|---|---|---|
| | **Benign** | | **Malicious** | | **Benign** | | **Malicious** | | **Benign** | | **Malicious** | |
| | #s | % | #s | % | #s | % | #s | % | #s | % | #s | % |
| T1557 | 0.0 | 0.00% | 0.4 | 0.00% | 0.0 | 0.00% | 0.0 | 0.00% | 0.0 | 0.00% | 0.2 | 0.00% |
| T1135 | 24.0 | 0.03% | 362.2 | 0.39% | 8.0 | 0.03% | 114.2 | 0.37% | 8.0 | 0.03% | 119.6 | 0.39% |
| T1124 | 61962.6 | 66.56% | 71892.6 | 77.23% | 20654.2 | 66.56% | 23935.8 | 77.14% | 20654.2 | 66.56% | 23976.2 | 77.27% |
| T1071 | 9.0 | 0.01% | 46.8 | 0.05% | 3.0 | 0.01% | 14.2 | 0.05% | 3.0 | 0.01% | 14.8 | 0.05% |
| T1590 | 2.4 | 0.00% | 0.2 | 0.00% | 0.8 | 0.00% | 0.0 | 0.00% | 0.8 | 0.00% | 0.0 | 0.00% |
| T1563 | 0.0 | 0.00% | 0.0 | 0.00% | 0.0 | 0.00% | 0.0 | 0.00% | 0.0 | 0.00% | 0.0 | 0.00% |
| T1105 | 97.2 | 0.10% | 42.4 | 0.05% | 32.4 | 0.10% | 12.4 | 0.04% | 32.4 | 0.10% | 11.4 | 0.04% |
| T1090 | 1.2 | 0.00% | 44.6 | 0.05% | 0.4 | 0.00% | 10.4 | 0.03% | 0.4 | 0.00% | 15.0 | 0.05% |
| T1550 | 0.6 | 0.00% | 10.8 | 0.01% | 0.2 | 0.00% | 3.6 | 0.01% | 0.2 | 0.00% | 3.6 | 0.01% |
| T1570 | 0.0 | 0.00% | 0.0 | 0.00% | 0.0 | 0.00% | 0.0 | 0.00% | 0.0 | 0.00% | 0.0 | 0.00% |
| T1571 | 71584.2 | 76.90% | 86088.8 | 92.48% | 23861.4 | 76.90% | 28652.6 | 92.34% | 23861.4 | 76.90% | 28683.8 | 92.44% |
| T1053 | 3.0 | 0.00% | 0.8 | 0.00% | 1.0 | 0.00% | 0.2 | 0.00% | 1.0 | 0.00% | 0.0 | 0.00% |
| T1021 | 11.4 | 0.01% | 28.6 | 0.03% | 3.8 | 0.01% | 8.4 | 0.03% | 3.8 | 0.01% | 8.6 | 0.03% |
| Tnull | 21548.0 | 23.15% | 6986.0 | 7.50% | 7045.0 | 22.70% | 2416.0 | 7.79% | 7178.0 | 23.13% | 2299.0 | 7.41% |
| **Total** | 133695.6 | | 158518.2 | | 44565.2 | | 52751.8 | | 44565.2 | | 52833.2 | |

Table 4 describes the mean statistics of our data across all five of our training, validation, and test sets. It also shows for each individual TTP:

- The number of malicious and benign samples that contain the TTP.
- The percentage of malicious and benign samples that contain the TTP.

When we look at the last row of Table 4 which shows the total number of samples that match a TTP (excluding samples that do not match any TTP), we observe that there isn't a significant difference between the number of benign and malicious samples with TTPs ($\sim$15% difference). Further, notice that the average number of TTPs matched by benign samples is 1.4, while it is 1.7 for malicious ones. The second observation is that the dataset is highly unbalanced at the TTP level: T1124 and T1571 occur in most of the malicious and benign samples, while most of the others occur rarely in the datasets, with T1557, T1590, T1563, T1570, T1053 not occurring at least once on average in our train/validation/test sets. This implies that our models cannot always be properly trained/validated/tested on the labels corresponding to such TTPs as there are no samples with such labels in the corresponding train/validation/test sets. For this reason, we do not consider the labels associated with T1557, T1590, T1563, T1570, T1053, and only consider the 9 labels T1135, T1124, T1071, T1105, T1090, T1550, T1571, T1021, each associated to the corresponding TTP, plus the label Mal-Sample corresponding to the entire sample. Finally, while the most frequent TTPs often occur in malicious samples, the opposite happens for some other TTPs, like T1105 and T1053.

## 4.2. Metrics

Given the $i$th label $A_i$ (representing either a TTP or the label Mal-Sample associated to the sample), we assume

1) $P_i$ and $N_i$ denote the sets of malicious and benign samples with label $A_i$ (and thus $|P_i| + |N_i| = n$ where $n$ is the number of samples), and
2) $PP_i$ and $PN_i$ denote the sets of predicted malicious and predicted benign samples respectively, with label $A_i$ (and thus also $|PP_i| + |PN_i| = n$).

Then, all the metrics are defined on the basis of the *confusion matrix* represented by the different values for $P_i$, $N_i$, $PP_i$ and $NN_i$ of the different labels. Considering the $i$th label $A_i$,

1) the *True Positive set* $TP_i$ is the set of samples which are predicted to be positive (malicious) and are actually positive (malicious):
$$TP_i = P_i \cap PP_i.$$
2) the *False Positive set* $FP_i$ is the set of samples which are predicted to be positive (malicious) and are actually negative (benign):
$$FP_i = N_i \cap PP_i.$$
3) the *True Negative set* $TN_i$ is the set of samples which are predicted to be negative (benign) and are actually negative (benign):
$$TN_i = N_i \cap PN_i.$$

TABLE 5: Average and standard deviation of the results obtained by running STTP2STTP and S2STTP on the five datasets, detecting the malicious TTPs. For each metric, the ↑ (resp. ↓) indicates that the higher (resp. lower) the result, the better the performance. For each pair of results, the best is in bold.

| | STTP2STTP | S2STTP |
|---|---|---|
| $F_1$-score (↑) | **0.835 ± 0.003** | 0.826 ± 0.003 |
| Accuracy (↑) | **0.972 ± 0.000** | 0.970 ± 0.000 |
| Subset Accuracy (↑) | **0.841 ± 0.002** | 0.831 ± 0.002 |
| Precision (↑) | **0.953 ± 0.002** | 0.951 ± 0.002 |
| Recall (↑) | **0.744 ± 0.005** | 0.730 ± 0.005 |
| FPR (↓) | **0.003 ± 0.002** | 0.004 ± 0.000 |

4) the *False Negative set* $FN_i$ is the set of samples which are predicted to be negative (benign) and are actually positive (malicious):

$$FN_i = P_i \cap PN_i.$$

On the basis of the confusion matrix of each label, many metrics have been defined and used to evaluate model performance in multi-label and binary classification problems. A standard way to present the different metrics is to distinguish between "micro" and "macro" average scores, the latter usually distinguished by appending the word "Macro" in front or after the name of the metric.

In the micro case, each metric groups the classifications by sample across all labels, and then calculates the overall metric, thus giving each prediction equal importance. Hence favouring the labels with high *support*, defined as the number of samples having a particular label. In contrast, each macro metric calculates the average of the metric across all classes, thus giving each label equal importance, irrespective of their support. In general, macro metrics are better suited in highlighting differences in performance when considering highly unbalanced datasets, as is our case when looking at the data at the TTP-level.

We use the following micro and macro scores to present our results. See, e.g., [24] for a more in depth presentation of such metrics, including a study about the correlations existing among them.

#### 4.2.1. Micro-average metrics.

If we define $TP = \sum_i |TP_i|$, $FP = \sum_i |FP_i|$, $TN = \sum_i |TN_i|$, $FN = \sum_i |FN_i|$, $P = \sum_i |P_i|$, $N = \sum_i |N_i|$ and $PP = \sum_i |PP_i|$ the following (micro-average) metrics are among the most commonly used:

1) *$F_1$-score ($F_1$)*, it is the harmonic mean of Precision and Recall, defined as:

$$F_1 = \frac{2PrRe}{Pr + Re} = \frac{2TP}{2TP + FP + FN} = \frac{2TP}{PP + P}.$$

2) *Accuracy (Acc)*, it is defined as the ratio between the number of correctly predicted labels to the total number of predictions:

$$Acc = \frac{TP + TN}{TP + TN + FP + FN} = \frac{TP + TN}{P + N}.$$

3) *Precision (Pr)*, it is the proportion of the correctly predicted positive labels, i.e., the ratio between the correctly predicted labels and the total number of predicted labels:

$$Pr = \frac{TP}{TP + FP} = \frac{TP}{PP}.$$

4) *Recall (Re)*, it is the ratio between the correctly predicted labels and the total number of actual labels:

$$Re = \frac{TP}{TP + FN} = \frac{TP}{P}.$$

5) *False Positive Rate (FPR)*, it is the proportion of the positive cases that were incorrectly identified:

$$FPR = \frac{FP}{FP + TN} = \frac{FP}{N}.$$

Another standard metric used in multi-label classification, defined directly on the basis of the ground truth $y_i$ and the predicted labels $\hat{y}_i$ of each sample $x_i$, is Subset Accuracy. *Subset Accuracy (SA)* is defined as the percentage of samples that have all their labels classified correctly:

$$SA = \frac{1}{n}\sum_{i=1}^{n} 1(y_i = \hat{y}_i)$$

where $n$ is the total number of samples and $1(.)$ is the indicator function returning 1 if the condition in parenthesis is true, and 0 otherwise.

#### 4.2.2. Macro-average metrics.

In multi-label classification problems, macro-average metrics are computed by averaging the corresponding micro-average metric as defined in Section 4.2.1. So, if $N$ is the number of labels, *Macro $F_1$-score* is defined as:

$$\textit{Macro } F_1\textit{-score} = \frac{1}{N}\sum_i \frac{2|TP_i|}{2|PP_i| + |P_i|},$$

*Macro Precision* as:

$$\textit{Macro Precision} = \frac{1}{N}\sum_i \frac{|TP_i|}{|PP_i|},$$

and *Macro Recall* as:

$$\textit{Macro Recall} = \frac{1}{N}\sum_i \frac{|TP_i|}{|P_i|}.$$

In our case $N = 9$, corresponding to the 8 TTPs appearing on average at least once in our train/validation/test sets, plus the label Mal-Sample associated to the entire sample.

### 4.3. Evaluating Malicious Usage of TTPs

Our first experiment tests the effectiveness of detecting malicious usage of TTPs and assesses the extent to which providing TTP features is helpful. We thus consider our STTP2STTP and S2STTP models with the following hyperparameters: 1250 hidden dimensions, dropout rate of 0.4, learning rate of $1.6 \times 10^{-4}$, and weight decay of $3.05 \times 10^{-6}$. All the models are trained using mini-batching (with a batch size of 750) and Adam optimizer [16]. The results for the different metrics are in Table 5, where the best results are highlighted in bold.

Looking at the table, we observe that, as expected, providing TTP features consistently helps across all metrics.

TABLE 6: Average of the scores corresponding to each label. For each metric, the ↑ (resp. ↓) indicates that the higher (resp. lower) the result, the better the performance. For each pair of results, the best is in bold. For each label, column Support reports the number of (malicious) samples with that label.

| | STTP2STTP | | | S2STTP | | | |
|---|---|---|---|---|---|---|---|
| | $F_1$ (↑) | Precision (↑) | Recall (↑) | $F_1$ (↑) | Precision (↑) | Recall (↑) | Support |
| T1135 | **0.962** | **0.986** | **0.940** | 0.872 | 0.884 | 0.866 | 119.6 |
| T1124 | **0.850** | **0.952** | **0.768** | 0.838 | 0.950 | 0.752 | 23976.2 |
| T1071 | **0.656** | **0.742** | **0.590** | 0.566 | 0.806 | 0.458 | 14.8 |
| T1105 | **0.686** | **0.660** | **0.726** | 0.000 | 0.000 | 0.000 | 11.4 |
| T1090 | **0.946** | **0.972** | **0.922** | 0.078 | 0.600 | 0.044 | 15.0 |
| T1550 | **0.160** | **0.200** | **0.134** | 0.000 | 0.000 | 0.000 | 3.6 |
| T1571 | **0.840** | **0.952** | **0.752** | 0.830 | 0.954 | 0.736 | 28683.8 |
| T1021 | 0.000 | 0.000 | 0.000 | **0.310** | **0.594** | **0.236** | 8.6 |
| Mal-Sample | **0.820** | **0.952** | **0.718** | 0.810 | 0.950 | 0.708 | 31031.0 |
| Macro avg | **0.658** | **0.713** | **0.617** | 0.478 | 0.638 | 0.422 | 9318.2 |

TABLE 7: Average and standard deviation of the results obtained by running each of the four models on the five datasets, to detect malicious samples. For each metric, the ↑ (resp. ↓) indicates that the higher (resp. lower) the result, the better the performance. For each metric, the best result is in bold, and the second best is underlined. The last row reports the average ranking.

| | STTP2STTP | S2STTP | STTP2S | S2S |
|---|---|---|---|---|
| $F_1$-score (↑) | **0.820 ± 0.002** | <u>0.811 ± 0.003</u> | <u>0.811 ± 0.002</u> | 0.802 ± 0.004 |
| Accuracy (↑) | **0.842 ± 0.002** | <u>0.835 ± 0.002</u> | 0.834 ± 0.001 | 0.827 ± 0.004 |
| Precision (↑) | **0.952 ± 0.002** | <u>0.951 ± 0.002</u> | 0.939 ± 0.005 | 0.949 ± 0.005 |
| Recall (↑) | **0.720 ± 0.004** | 0.708 ± 0.004 | <u>0.715 ± 0.004</u> | 0.700 ± 0.002 |
| FPR (↓) | **0.037 ± 0.002** | **0.037 ± 0.002** | <u>0.047 ± 0.004</u> | 0.045 ± 0.006 |
| Avg ranking | 1.1 | 2.2 | 3.1 | 3.6 |

Secondly, the absolute value is very positive for each metric, and in particular, it is remarkably high for Accuracy ($\simeq 0.97$, meaning that more than 97% of the predictions are correct) and Precision ($\simeq 0.95$, meaning that more than 95% of the positive predictions are correct). We obtain the "lowest" value for the Recall metric ($\simeq 0.74$, meaning that our models are able to detect roughly three out of four of the positive labels, each representing maliciousness). Next, we see that though the TTP information helps, the difference between the values associated with each metric does not appear to be remarkably high if we consider the (micro) statistics in Table 5. This fact underlines that, assuming we are interested in maximising such micro-average scores, it is possible to not provide the TTP features as input and let the model learn how to detect and properly classify each TTP by itself, with a relatively low decay in the micro-average performance. However, a careful analysis of the Precision, Recall, and $F_1$-score of each label, reported in Table 6, reveals that:

1) TTP features are beneficial in all cases, except for T1021.
2) The contribution of the TTP features is relatively modest for the highly occurring TTPs (T1124 and T1571), and significantly higher for many of the others. For instance, if we consider T1105, the $F_1$-score for STTP2STTP is 0.686, and 0 for S2STTP.

Such facts are reflected by the macro-average scores reported in the last line of Table 6, where we see huge improvements across all metrics:

1) Macro $F_1$-score jumps from 0.478 to 0.658, with an increase of 37.55%.
2) Macro Precision goes from 0.638 to 0.713, with an increase of 11.82%.
3) Macro Recall jumps to 0.617 from 0.422, with the highest increase of 46.05%.

In Section 5.1 we will see how such positive results can further be improved by fine-tuning the threshold $\theta$ associated with each label to decide the maliciousness of the corresponding TTP and/or sample.

As a final consideration, we remark that all the statistics reported in Tables 5 and Table 6 take into account the metrics associated with all the labels, including the label Mal-Sample associated with the entire sample. This is standard practice in both multi-label classification problems and hierarchical multi-label classification problems (see, e.g., [39]). Further, if we do not count the outputs of Mal-Sample, the statistics in Table 5 and the macro-average statistics in Table 6 would only be moderately affected. Indeed, the Precision and Recall of the label Mal-Sample associated with the sample (in Table 6) are very close to the overall Precision and Recall in Table 5, and consequently the two $F_1$-scores of Mal-Sample computed by STTP2STTP and S2STTP (equal to $\simeq 0.82$ and $\simeq 0.81$, respectively) are also similar to the overall $F_1$-score in Table 5 (equal to $\simeq 0.83$).

## 4.4. Evaluating Sample Detection

We now want to evaluate how the detection of TTPs and their classification as malicious has an impact on the classification of the entire sample as either malicious or benign, and the impact of providing TTP features. We implement STTP2S and S2S with the same configuration and hyper-parameters as STTP2STTP and S2STTP.

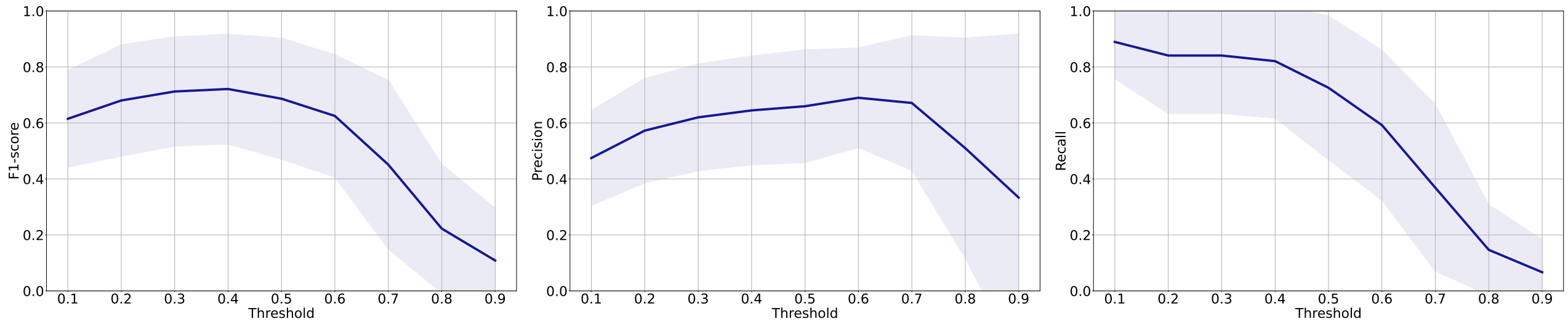

Figure 1: Average and confidence intervals of $F_1$-score, Precision and Recall for TTP T1105 while varying the threshold $\theta$ of the TTP label.

TABLE 8: Precision, Recall and $F_1$-score at different thresholds with the STTP2STTP model. $\theta_{max}$ is the threshold in between [0.1,0.9], step 0.1, for which we get the maximum resulting value. For $\theta = 0.5$ we report the same values of Table 5 in order to make comparison easier.

| | $F_1$ score | | | | | Precision | | | | | Recall | | | | |
|---|---|---|---|---|---|---|---|---|---|---|---|---|---|---|---|
| | $\theta$ | value | $\theta_{max}$ | value | inc. | $\theta$ | value | $\theta_{max}$ | value | inc. | $\theta$ | value | $\theta_{max}$ | value | inc. |
| T1135: | 0.5 | 0.963 | 0.4 | **0.967** | 0.4% | 0.5 | 0.988 | 0.8 | **0.994** | 0.7% | 0.5 | 0.940 | 0.1 | **0.982** | 4.4% |
| T1124: | 0.5 | 0.850 | 0.4 | **0.851** | 0.1% | 0.5 | 0.952 | 0.9 | **0.979** | 2.9% | 0.5 | 0.768 | 0.1 | **0.998** | 30.1% |
| T1071: | 0.5 | 0.656 | 0.3 | **0.716** | 9.2% | 0.5 | 0.741 | 0.3 | **0.891** | 20.3% | 0.5 | 0.591 | 0.1 | **0.820** | 38.7% |
| T1105: | 0.5 | 0.687 | 0.4 | **0.721** | 5.0% | 0.5 | 0.660 | 0.6 | **0.690** | 4.6% | 0.5 | 0.726 | 0.1 | **0.890** | 22.6% |
| T1090: | 0.5 | 0.947 | 0.6 | **0.953** | 0.7% | 0.5 | 0.973 | 0.6 | **0.987** | 1.4% | 0.5 | 0.922 | 0.1 | **0.946** | 2.6% |
| T1550: | 0.5 | 0.160 | 0.1 | **0.652** | 307.7% | 0.5 | 0.200 | 0.1 | **0.610** | 205.0% | 0.5 | 0.133 | 0.1 | **0.771** | 478.6% |
| T1571: | 0.5 | 0.841 | 0.4 | **0.842** | 0.2% | 0.5 | 0.954 | 0.9 | **0.980** | 2.8% | 0.5 | 0.752 | 0.1 | **0.998** | 32.8% |
| T1021: | 0.5 | 0.000 | 0.1 | **0.141** | | 0.5 | 0.000 | 0.1 | **0.078** | | 0.5 | 0.000 | 0.1 | **0.793** | |
| Mal-Sample | 0.5 | 0.820 | 0.4 | **0.821** | 0.2% | 0.5 | 0.952 | 0.9 | **0.980** | 2.9% | 0.5 | 0.719 | 0.1 | **0.998** | 38.7% |
| Macro avg | | 0.658 | | **0.741** | 12.5% | | 0.713 | | **0.799** | 12.0% | | 0.617 | | **0.911** | 47.6% |

The results obtained by running all the models (i.e., STTP2S and S2S as well) are reported in Table 7, where the best results are highlighted in bold, while the second best are underlined. The last line reports the average ranking, obtained by ranking the performance of each system on a metric and then averaging the results. In the case of ties between models, average ranks are assigned to such models (see [6] for more details).

Considering the table, the first observation is again that STTP2STTP has the best performance across all the different metrics. Further, STTP2STTP almost always has the most stable performance of the lot, as measured by the standard deviation: the only exception to the rule is for Recall, for which STTP2STTP has a standard deviation of 0.004, equal to the standard deviation of S2STTP and STTP2S, but higher than that of S2S. Then, the second most performing model is S2STTP, which underlines the fact that the precise detection and classification of TTPs improves the overall performance of the corresponding binary classifiers. This result may look surprising given that $(i)$ both STTP2STTP (resp. S2STTP) has been trained on the same dataset used for training STTP2S (resp. S2S), and $(ii)$ STTP2STTP and S2STTP are actually solving a more complex problem than their corresponding models STTP2S and S2S, namely identifying malicious usage of TTPs and determining the maliciousness of the overall sample. However, by having a separate label for each TTP, the neural network has now the possibility to focus on the detection of the malicious samples containing rarely used TTPs, and classify them accordingly. The relatively good performance of S2STTP even when used as a binary classifier confirms that $(i)$ it is possible to not provide TTP features as input, $(ii)$ have the model learn how to detect the malicious usage of TTPs, and $(iii)$ still not get a huge performance decay.

Additionally, model STTP2S has better performance than of S2S, confirming the fact that providing TTP information improves the model's performance, as also highlighted by the better performance of STTP2STTP compared to S2STTP.

Overall, the main results of this experimental analysis are that:

1) Teaching the neural network to detect the malicious usage of TTPs helps to improve the performance, as demonstrated by the good performances obtained by STTP2STTP and S2STTP,
2) Providing TTP features improves performance, as demonstrated by the performance of STTP2STTP vs S2STTP and STTP2S vs S2S, and
3) Teaching the neural network to detect the malicious usage of TTPs and providing TTP features, leads to the best overall results, as demonstrated by STTP2STTP and its results across all metrics.

## 5. Experimental Analysis

In this Section, we present the results of further experimental analysis showing the results of $(i)$ tuning the label decision thresholds in STTP2STTP (Subsection 5.1), $(ii)$ training STTP2STTP and S2STTP with fewer samples (Subsection 5.2), and $(iii)$ testing STTP2STTP and S2STTP with samples in which benign flows have been injected in the malicious samples in an attempt to camouflage their malicious activity (Subsection 5.3).

## 5.1. Threshold Variation per TTP

The already positive performances of STTP2STTP reported in Section 4.3, can be further improved by tuning the decision threshold $\theta$ associated to each label and above which the corresponding TTP is predicted to have been maliciously used. For instance, if we consider TTP T1105 and plot its $F_1$-score, Precision and Recall scores while varying its threshold, we obtain the graphs in Figure 1. As it can be observed, T1105 $F_1$-score peaks at some value in between $(0, 1)$ which is not necessarily equal to $0.5$, which is the standard and intuitive threshold we used in our experiments. Indeed, we can expect $F_1$-scores to have a low value when the threshold is (extremely) low or (extremely) high. For instance, considering T1105, the average $F_1$-score across the five datasets is equal to $11.4/62061 \simeq 2 \times 10^{-4}$ when $\theta = 0$ (i.e., equal to the ratio between its average support, and the average number of samples in the test set), and equal to 0 when $\theta = 1$ (as in this case no sample is predicted to maliciously use T1105). At the standard setting $\theta = 0.5$, the $F_1$-score for T1105 is 0.686, which raises to 0.721 (5.1% improvement) if $\theta$ is set to 0.4. A similar behaviour can be observed for T1105 Precision, which peaks at 0.6. Indeed, for $\theta = 0$, every sample is positively predicted and thus the Precision is equal to the ratio between the support of the label and the number of samples in the test set, which for T1105 is equal to $11.4/62061 \simeq 2 \times 10^{-4}$ on average. Then, we can expect the Precision to increase with $\theta$ under the assumption that the value returned by the neural network reflects the probability of the TTP being maliciously used. However, as the figure shows, it may not be the case that Precision is ensured to be monotonically increasing as, in the limit case of $\theta = 1$, no labels are predicted and thus the Precision gets the undefined value $0/0$. On the contrary, Recall is bound to monotonically decrease from value 1 for $\theta = 0$ to value 0 for $\theta = 1$, as it is indeed the case for the Recall of T1105. All such considerations are reflected by the plots in Figure 1, which also shows the confidence interval of each metric that, for Precision, gets relatively high as $\theta$ approaches the value 1.

Table 8 reports the best result for Precision, Recall, and $F_1$-score for the various labels while varying their respective thresholds. As it can be observed, substantial improvements can be obtained for either Precision or Recall or $F_1$-score, which one depending on the particular metric we are interested in for each single TTP. The substantial improvements are highlighted by the macro-average metrics reported in the last line:

1) Macro $F_1$-score goes from 0.658 to 0.741, with a 12.5% increase,
2) Macro Precision goes from 0.713 to 0.799, with a 12.0% improvement, and
3) Macro Recall jumps to 0.911 from 0.617, with the highest 47.6% improvement.

Of course, such a fine-tuning at the TTP level is also possible for S2STTP, while it is not for STTP2S and S2S.

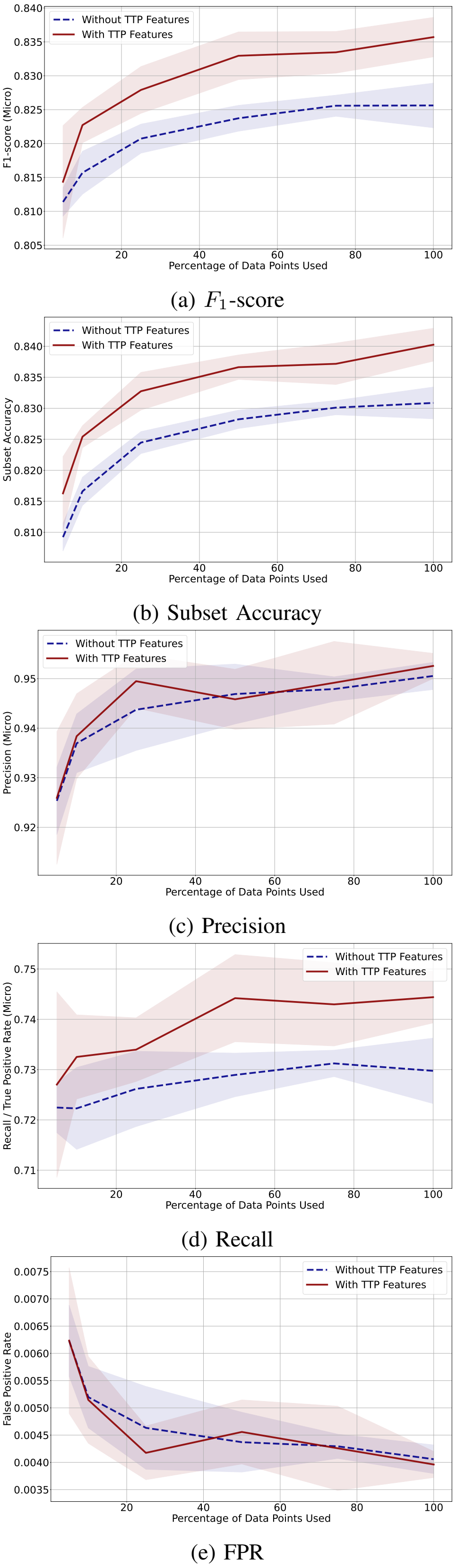

Figure 2: Performance of the models when varying the size of the training set. On the $x$-axis the percentage of randomly sampled datapoints from the training set can be found.

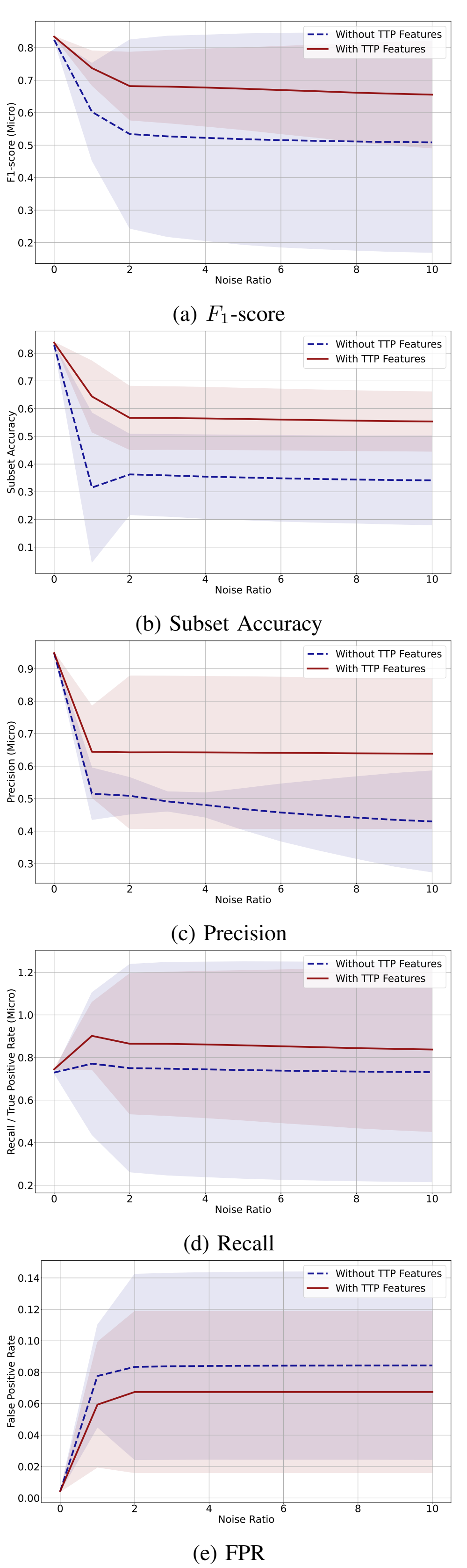

(a) $F_1$-score

(b) Subset Accuracy

(c) Precision

(d) Recall

(e) FPR

Figure 3: Performance of STTP2STTP and S2STTP when tested with varying percentages the noise ratio, defined as the ratio between the number of injected benign flows and the number of flows in the original sample.

## 5.2. Data Scarcity

One of the known problems of machine learning models is that their performance is very sensitive to the amount of data used for training. However, we expect the model STTP2STTP to be less dependent on the amount of data used for training, thanks to the injection of the TTP information. Thus, the existing gap between the performance obtained with STTP2STTP and S2STTP should either remain unvaried or increase as we train with less data. To test our hypothesis, we consider each of the five datasets used for training our models and randomly sample it in order to obtain 5 new datasets, each containing either 5%, or 10%, or 25%, or 50% or 75% of its datapoints. Thus, for each percentage, we have 5 new sets of samples, that we use to train both STTP2STTP and S2STTP, which are then tested on the same test sets used in Section 4.

The results obtained according to different metrics are shown in Figure 2, in which we also consider the performance of the models when trained with 100% of the samples. We also plot the 95% confidence interval for each result, which is represented by the shading around each line in the figure.

Considering the figures, we see that:

1) as expected, STTP2STTP has better performance than S2STTP when considering $F_1$-score, Subset Accuracy and Recall,
2) even for such metrics the gap between the two models varies depending on the percentage of data points used for training, while
3) for Precision and False Positive Rate, for some percentages, S2STTP has better performance than STTP2STTP, though this happens only for very few points.

Such behaviours can be explained by considering again the average characteristics of our five datasets reported in Table 4. Indeed, the labels with low support in the original datasets will very likely have correspondingly even lower support in the new sets used for training, as the percentage used for sampling decreases. This likely corresponds to a corresponding degradation of the performance, which is (almost always) reflected by the plot in the figure, and which we expect to be particularly the case for the labels with low support. At the same time, the micro-average scores in the Figure, favouring the highly supported labels, will cause the neural network to reward the still good performance it can get (even when training on 5% of the datasets) for the labels T1124, T1571 and Mal-Sample. Indeed, when sampling with the 5% percentage,

1) T1124, T1571 and Mal-Sample are expected to have, on average, 3594.63, 4304.44 and 4654.65 as support in the dataset used for training,
2) all the other ones (except for T1135 which still should have a support $\simeq$ 18.11) should have support smaller than 3, and
3) the micro-average scores of all the labels different from T1124, T1571 and Mal-Sample are in the vast majority of cases equal to 0, even when given the TTP information.

Thus, in many cases, when considering the performance of STTP2STTP and S2STTP trained on smaller datasets, they are mostly considering just the labels T1124, T1571, and Mal-Sample, each of them with relatively high support in the training set even when considering 5% of the samples, thus allowing also S2STTP to recognise the malicious usage of the TTPs and the sample maliciousness.

## 5.3. Adversarial Benign Flow Injection

In this last experiment, we consider the scenario in which a malicious attacker tries to camouflage his malicious activity by doing various other benign activities within the same traffic sample. To test the robustness of both STTP2STTP and STTP2S to such scenario,

1) we consider benign traffic flows extracted from the Ember Benign dataset in Table 3,
2) to each malicious sample, we add a varying percentage of benign flows, starting from 0% up to 1000% with step 100%, and
3) we test STTP2STTP and S2STTP performance without retraining such systems.

We measure the activity of the attacker in camouflaging its malicious activity in terms of the Sample Noise Ratio, defined as the ratio of injected benign flows to malicious flows (thus, a sample noise ratio equal to 2 means that each malicious sample with $f$ flows has been added with $2f$ benign flows). Indeed, we expect that on such new test sets, both STTP2STTP and S2STTP should have a decay in performance given that the newly generated samples have different characteristics from the ones of the samples used for training both systems. The research questions we want to investigate are $(i)$ whether the TTP information still helps STTP2STTP in detecting the malicious usage of the TTPs, $(ii)$ to what extent STTP2STTP and S2STTP performance decay, and $(iii)$ whether the performance gap between STTP2STTP and S2STTP increases or decreases with the sample noise ratio.

Figure 3 shows the plots for the $F_1$-score, Subset Accuracy, Precision, Recall, and False Positive Rate, in their micro-average form. On the $x$-axis there is the used sample noise ratio.

As it can be observed from the figures, the results confirm our expectations:

1) STTP2STTP has always better performance than S2STTP,
2) the performance gap between STTP2STTP and S2STTP increases with a rather steep gradient up to a certain point after which the difference becomes roughly constant or only marginally increasing with the sample noise ration, and
3) both models have a large standard deviation, usually S2STTP larger than STTP2STTP with the exception for Recall for which the opposite happens.

As we said, such behaviours were somehow expected. The only relatively surprising fact is that the scores tend to stabilise when the number of injected flows is 200%. However, this fact can be easily explained by looking at the plot for Subset Accuracy, which measures the number of samples that are exactly labelled by the models. As it can be seen, from noise ratio 2 on, both STTP2STTP and S2STTP have an (almost) constant Subset Accuracy, implying that in the original testing set there are:

1) malicious samples whose traffic flow (even when represented as a bag of flow) still make the detection of the used TTP robust to the amount of noise added, even when not using the TTP information, and
2) the number of such samples becomes bigger when using TTP information.

Such analysis is confirmed by computing the macro-average metrics which have exactly the same behaviour of the micro-average metrics in Figure 3.

# 6. Discussion

**Consistent Improvement over the State-of-the-Art.** Our results highlight the consistency of our system's improvement over the state-of-the-art when it comes to detecting malware with the help of TTP features. Both versions of our system based on our multi-label DNN architecture (STTP2STTP and S2STTP) clearly outperform the binary DNN systems (STTP2S and S2S). This further highlights the power of our novel multi-label DNN architecture, since it enables us to outperform models like STTP2S that utilise TTP features, even when using a model like S2STTP which does not utilise TTP features.

**Explainability and Precision in Detecting Malware.** Our overall malware detection is not only better but also much more precise, as a result of our system being able to identify the malicious usage of each TTP individually. Instead of a binary output indicating the malicious or benign nature of a sample, our system can additionally identify which particular behaviours, as characterised by specific TTPs, were responsible for a malicious classification. Thus, making the output not only precise, but also more explainable to the user. For instance, the mean Accuracy of STTP2STTP over all our datasets increases by 0.83%, 0.95%, and 1.81% when compared to S2STTP, STTP2S, and S2S.

**Recognition of rarely-occurring TTPs.** ML-based malware detection systems rely on ample amounts of representative training data in order to reliably detect similar or perhaps unseen malware. Similarly, ML-based systems that rely on TTPs also require training on a significant number of datapoints for each TTP in order to later identify them. As a result, unlike other ML-based systems where TTPs with only a few datapoints might be overlooked during model training, our system can also accurately identify malicious behaviour using TTPs which only have a few datapoints present in the training data. As shown in our experiments, our system can identify malware using TTPs with as few as 15 datapoints with an average $F_1$-score of 0.946, as in the case of T1090.

**System Configurability.** A key part of our system design is that it can detect malicious usage of individual TTPs, thus allowing a user to choose the specific TTPs most relevant to their threat model. Additionally, we demonstrate that we can tune the detection thresholds for each TTP in our system, to enable even greater granularity of detection based on specific requirements. For instance, the threshold for a rarely-occurring but high-risk TTP can be lowered in a given environment, while a commonly-utilised TTP which does not pose a significant risk can have its detection threshold increased, thus increasing and reducing the number of positively detected samples respectively. Our results across our datasets highlight the benefits of this configurability—resulting in an average improvement of Macro F1-score by 12.5%, Macro Precision by 12%, and Macro Recall by 47.6%.

**Utility of TTPs in Malware Detection.** Our experimental analysis highlights the utility of TTP information in improving malware detection capabilities. When utilising our system with TTPs (STTP2STTP), we see an average $F_1$-score improvement of 1.08% at the sample level and 37.65% at the TTP-level, as compared to our system without TTPs (S2STTP). This can further be seen when comparing results at just the sample level (i.e., binary classification)—on average STTP2STTP outperforms S2STTP, STTP2S, and S2S across every metric ($F_1$-score, Accuracy, Precision, Recall, and FPR)—thus showing the utility of TTPs in improving standard malware detection capabilities.

**Data Scarcity.** We find that even when utilising limited training data, the addition of TTP features in our system architecture helps improve overall performance. Across the 30 experiments we performed (5 datasets x 6 training data percentages) STTP2STTP outperforms S2STTP 96.67% of the time according to the $F_1$-score. The addition of TTPs significantly improves the capability of the system even when using much less training data. For instance, STTP2STTP utilising 25% of the training data has a higher average $F_1$-score than S2STTP utilising 100% of the training data, and STTP2STTP utilising 10% of the training data has a higher average Recall than S2STTP utilising 100% of the training data.

**Resilience to Adversarial Attack.** Our experimental analysis also evaluates the capability of our system to resist adversarial attack, specifically camouflaging attacks, where an adversary deliberately injects benign flows into a malicious sample in order to camouflage its malicious activity. We measure this in terms of the Sample Noise Ratio (ratio of injected benign flows to malicious flows), and find that while system performance decreases as expected when the Noise Ratio increases, STTP2STTP is consistently able to perform better under these conditions than S2STTP. Indeed, from an average improvement of 1.21% in $F_1$-score when the Noise Ratio is 0.0, we see an average improvement of up to 28.93% in $F_1$-score when the Noise Ratio is 10.0. Similarly, we observe a consistent average improvement across every metric in the noisiest situation—62.17% in Subset Accuracy, 48.71% in Precision, 14.50% in Recall, and 20.23% in terms of FPR.

**Minimal Additional Data.** Our approach also highlights the minimal data requirements for achieving these significant performance gains. The TTPs we utilise in our system, as described in the MITRE ATT&CK framework, are open-source information, and our procedure to create TTP features using them does not require additional training data. Since these features can be created directly from existing data, utilising open-source domain knowledge regarding adversarial behaviour, it provides a vital performance benefit for the data-greedy machine learning algorithms.

**Different Levels of Granularity.** Our approach allows us to report the maliciousness of each sample at two levels of granularity which, as explained in [3], allows for different fine-grained outputs that can be used under divergent levels of expertise. In our case, we have just two levels (sample and TTPs) but it is of course possible to $(i)$ group the TTPs together according to the tactical goal of the attacker, $(ii)$ subdivide each technique into the various sub-techniques it may include, and $(iii)$ also include the procedures used to implement the technique or sub-technique. This would correspond to a five-level hierarchy reflecting the TTP ontology described in the MITRE ATT&CK framework, which, if incorporated into our model (thus having one hierarchical label per tactic, technique and procedure, plus the label for sample maliciousness) would give our system the ability to $(i)$ report a more fine-grained output, and $(ii)$ recognise a malicious activity at different levels of granularity, each with its own level of confidence. For instance, our system would be able to recognise a sample as malicious, e.g., at the tactical level while failing to recognise the specific technique and sub-technique or procedure being used: though not optimal, this would still be better than just reporting that the sample is malicious, or incorrectly detect the malicious usage of a specific technique and/or sub-technique and/or procedure.

## 7. Conclusions

In this paper, we have designed a novel multi-label DNN architecture that detects malicious behaviour at two different levels of granularity—for the TTPs and the overall sample. This unique ability allows us to improve the overall performance of malware detection, improve on the state-of-the-art when it comes to detecting malicious behaviour when exploiting TTPs, and identifying hard-to-detect malware which utilises uncommon or rarely-used TTPs. We back up our findings with an extensive comparative analysis, using a dataset of over 1.5 million real-world malware and benign samples. Lastly, we test the robustness of our system in various challenging conditions, such as when relying on limited training data or subjected to adversarial camouflage attacks, and demonstrate that it continues to outperform other comparable systems.

As discussed, our system requires a minimal amount of additional data when compared to traditional models for detecting malicious activities and is able to provide precise outputs about which TTPs being used maliciously used. This opens up the possibility of future extensions in which the model will have a richer hierarchical representation of the TTPs, thus allowing for a finer level of granularity when it comes to the malware detection.

## Data Availability

All the materials necessary to reproduce this work, including source code and data, will be made available once the paper is published.